\documentclass[twocolumn,aip,floats]{revtex4}
\usepackage{graphicx}
\usepackage{amssymb}

\newcommand{\vect}[1]{{\mbox{\boldmath $#1$}}}

\begin{document}

\title{Gilbert Damping in Magnetic Multilayers}


\author{E. \v{S}im\'{a}nek}
 \email{simanek@onemain.com}

\affiliation{6255 Charing Lane, Cambria, CA 93428, USA}
\author{B. Heinrich}
\affiliation{Simon Fraser University, 8888 University Dr.,
Burnaby, BC, V5A 1S6, Canada}

\begin{abstract}
We study the enhancement of the ferromagnetic relaxation rate in
thin films due to the adjacent normal metal layers. Using linear
response theory, we derive the dissipative torque produced by the
$s-d$ exchange interaction at the ferromagnet-normal metal
interface. For a slow precession, the enhancement of Gilbert
damping constant is proportional to the square of the $s-d$
exchange constant times the zero-frequency limit of the frequency
derivative of the local dynamic spin susceptibility of the normal
metal at the interface. Electron-electron interactions increase
the relaxation rate by the Stoner factor squared. We attribute the
large anisotropic enhancements of the relaxation rate observed
recently in multilayers containing palladium to this mechanism.
For free electrons, the present theory compares favorably with
recent spin-pumping result of Tserkovnyak et al. [Phys. Rev. Lett.
\textbf{88},117601 (2002)].

\end{abstract}

\maketitle

\section{Introduction}
Ferromagnetic multilayers have attracted much attention recently
because of their applications in spintronics and high density
magnetic recording devices.  The present paper is concerned with
magnetic relaxation in a ferromagnetic film ($F$) imbedded between
nonmagnetic metallic layers ($N$). In particular, we study the
enhancement of the Gilbert damping in $N/F/N$ sandwiches as
compared with that of a single ferromagnetic film.  The Gilbert
damping constant $ G$ is defined by the Landau-Lifshitz-Gilbert
(LLG) equation of motion [1,2]

\begin{equation}\label{LL}
  \frac{1}{|\gamma|}\frac{\partial \vect{M}}{\partial t} =
  - \left[ \vect{M} \times \vect{H}_{eff} \right] +
  \frac{G}{\gamma^2 M_{s}^{2}} \left( \vect{M}\times\frac{\partial \vect{M}}
  {\partial t} \right) \; ,
\end{equation}
where $\vect{M}$ is the magnetization vector, $M_{s}$ is the
saturation magnetization, and $\vect{H}_{eff}$ is the effective
field which is given by $\vect{H}_{eff}=-\partial E/\partial \vect
{M}$ where $E$ is the Gibbs free energy. The gyromagnetic ratio
$\gamma$ is a negative quantity $-g\mu_B/\hbar$ where $g$ is the
spectroscopic splitting factor, and $\mu_B$ is the Bohr magneton.
The second term on the R. H. S. of Eq.(1) represents the
dissipative torque, the magnitude of which is proportional to $G$.
The LLG equation describes well both the static and the dynamic
properties of ultrathin ferromagnetic films and the $N/F/N$
sandwiches (for a review see Ref.[3]). The experimental and
theoretical aspects of spin relaxation in multilayers are covered
in Ref.[4].

Recent studies of the FMR linewidth for ultrathin films [5,6] show
that the constant $G$ is enhanced when a nonmagnetic metal is
deposited on the ferromagnetic film. This effect has been
predicted in 1996 by Berger [7] in his study of spin transfer
between the polarized conduction electrons and the ferromagnetic
film. Berger shows that there is an enhanced electron-magnon
scattering caused by isotropic $s-d$ exchange taking place at the
$F/N$ interface. Being a surface effect, the enhancement of the
Gilbert constant is inversely proportional to the thickness of the
ferromagnetic film. This unique feature is observed in recent
measurements of the additional FMR linewith on permalloy-normal
metal sandwiches [8], and on the double ferromagnetic layer
structures as well [9].  It should be mentioned that already in
1987 a study of FMR linewidth showed an appreciable increase in
the Gilbert damping with decreasing thickness of the $Fe$
ultrathin films grown on bulk $Ag(001)$ substrates [10].

A novel mechanism for additional Gilbert damping in $N/F/N$
structures has been recently proposed by Tserkovnyak et al.[11].
These authors calculate the spin current pumped through the $N-F$
contact by the precession of the magnetization vector
$\vect{M}(t)$.  The theory is based on extending the scattering
approach of parametric charge pumping by Brouwer [12] to spin
pumping.

Like the theory of Berger [7], the additional damping of Ref.[11]
scales inversely with the thickness of the ferromagnetic film
indicating that only the $F/N$ interface is involved. However, the
expression for excess $G$, which we call $G'$, differs
considerably from that of Ref.[7].  In particular, $ G'$ vanishes
with vanishing $s-d$ exchange splitting.  An attractive feature of
this theory is that it links $G'$ to the transport properties of
the interface. Due to exchange polarization of the $F/N$ contact,
the reflection $(r)$ and transmission $(t)$ coefficients at the
interface depend on the orientation of the conduction electron
spin with respect to the magnetization direction of the
ferromagnet.  The formula for $G'$ involves differences such as
$\Delta r=r^{\uparrow} - r^{\downarrow}$. Interestingly, similar
quantities play a role in the theory of interlayer magnetic
coupling by Bruno [13].  For instance, the
Ruderman-Kittel-Kasuya-Yosida (RKKY) coupling [14] between
two-dimensional layers can be obtained by calculating the
interlayer coupling energy in terms of the reflection coefficients
$r^{\uparrow}$ and $r^{\downarrow}$ of the layers.  These
coefficients are obtained by solving  a simple problem of
scattering by  $\delta$-function potential.  In the limit of weak
exchange splitting (compared to the Fermi energy), the derived
interlayer coupling agrees with the RKKY result of Yafet [15].

These considerations prompt us to take another look at the theory
of enhanced relaxation in multilayer.  The fact that the RKKY
theory [13] involves transport properties at the interface similar
to the spin pump theory of $G'$ [11] suggests that a suitable
generalization of RKKY theory to time-dependent magnetization
$\vect{M}(t)$ may unravel the needed dissipative torque of Eq.(1).
We note that the standard approach to RKKY coupling is to use a
linear response theory [14,16], and calculate the conduction
electron spin density induced by the contact exchange potential.
In applications to interlayer coupling between ferromagnetic
layers with time-independent magnetization vectors, it is the
static spin susceptibility of the electron gas which determined
the coupling.  In the present paper, we consider a response to a
slowly varying time-dependent $s-d$ exchange potential.

Owing to the dissipative part of the spin susceptibility, the spin
density induced by the precession of $\vect{M}(t)$ will have a
component that is out of phase with $\vect{M}(t)$.  Hence, such a
component will have a time-dependence given by $d \vect{M}/dt$.
The induced spin density has both local and nonlocal consequences
for the dynamics of multilayers.

First, let us focus on the local effect in the $N/F/N$ system. It
is instructive to invoke the analogy to radiation damping of
charged particle in classical electrodynamics [17]. Thus, we put
the dissipative torque of Eq.(1) in correspondence with the
radiation reaction force acting on the particle.  The first term
of Eq.(1) corresponds to the external force.  In view of this
analogy, we rewrite the dissipative torque in terms of a reaction
field $\vect{H}_{r}$

\begin{equation}\label{Eq2}
\frac{G'}{\gamma^{2} M^{2}_{s}}[\vect{M}\times
\frac{d\vect{M}}{dt}] = -[\vect{M}\times \vect{H}^{(d)}_{r}]
\end{equation}

where $\vect{H}^{(d)}_{r}$ represents the dissipative part of
$\vect{H}_{r}$. For the $N/F/N$ system shown in Fig.1., we find a
reaction field (see Appendix A)

\begin{equation}\label{Eq3}
\vect{H}_{r}(t) = \frac{2Ja}{\gamma
d}\langle\vect{s}(x=0,t)\rangle
\end{equation}

where $J$ is the $s-d$ exchange coupling constant, $a$ is of order
of the lattice constant, and $d$ is the width of the ferromagnetic
film. The quantity $\langle\vect{s}(x=0,t)\rangle$ is the spin
density induced at the $F/N$ interface by the $s-d$ exchange
interaction ($x$ being the distance from the interface). The
expression (3) is quite general in the sense that it can be used
for both ballistic and diffusive cases.  In the present paper we
confine ourselves to the ballistic case and calculate the induced
spin density using linear response theory [16].

Assuming slow precession of $\vect{M}(t)$, we find two
contributions:  One that is proportional to $\vect{M}(t)$ with a
coefficient given by the real part of the local spin
susceptibility at zero frequency.  If the spin susceptibility is
anisotropic, this term leads to an anisotropic shift of the FMR
frequency.  For isotropic susceptibility, the quantity on the
R.H.S. of Eq.(3) is a vector parallel to $\vect{M}(t)$ and the
corresponding torque (2) vanishes.

The second contribution to $\langle\vect{s}(x=0,t)\rangle$ is
proportional to $d\vect{M}/dt$.  The coefficient of
proportionality is the frequency derivative of the imaginary part
of the susceptibility taken at $\omega=0$.  Like the real part,
this quantity is generally anisotropic.  According to Eq.(2), it
produces a dissipative torque leading to an anisotropic $G'$.
Explicit evaluation of this term for an isotropic noninteracting
electron gas yields a formula for $G'$ which compares favorably
with the spin pump theory of Tserkovnyak et al.[11].

The proposed formulation allows us to incorporate interactions
between the electrons in the $N$ region.  The generalized
Hartree-Fock approximation shows that the $\vect{q},\omega$
dependent spin susceptibility of the interacting electron gas is
enhanced compared to the free electron case [16].  Using these
results, we find that the part of $\vect{H}_r$ that contributes to
the FMR frequency shift is enhanced by the Stoner factor
\begin{equation}\label{Eq4}
S_E=[1-U N (\epsilon_{F})]^{-1}
\end{equation}

where $U$ is the screened intraatomic Coulomb interaction and
$N(\epsilon_{F})$ is the electron density of states, per atom, at
the Fermi level [16].  On the other hand, we find that $G'$ is
enhanced by a factor of $S_E^{2}$.  The Stoner enhancement is
thought to be large in metals such as palladium and its alloys.
Recent results using multilayers with $Pd$ layers as a spacer show
a significant enhancement of interface damping exhibiting a
fourfold anisotropy in keeping with the present theory [18].

Let us turn to the nonlocal consequences of the dynamically
induced spin density.  This is relevant in the case of two
time-dependent magnetizations $\vect{M}_{1}(t)$ and
$\vect{M}_{2}(t)$ separated by a nonmagnetic spacer.  Similar to
the local effect, the magnetization $\vect{M}_{1}(t)$ at $x_{1}$
induces at the position $x_{2}$ of $\vect{M}_{2}(t)$ a spin
density oscillation with both the in phase
$(\varpropto\vect{M}_{1}(t))$ and the out of phase component
$(\varpropto d\vect{M}_{1}/dt)$. The interaction of the in-phase
component with $\vect{M}_{2}(t)$ yields the standard interlayer
RKKY coupling [15].  The out of phase component leads to a
dissipative interaction of the form $d\vect{M}_{1}(t)/dt
.\vect{M}_{2}(t)f(x_{2}-x_{1})$.  A similar form, with the indices
$1$ and $2$ interchanged, follows by taking into account the spin
density induced at $x_{1}$ by precession of $\vect{M}_{2}(t)$.
Such terms may be of importance in FMR studies involving
dynamically coupled magnetic films. However, these topics are
beyond the scope of the present investigation which focuses on
dynamics of single ferromagnetic film in contact with nonmagnetic
metal.

The paper is organized as follows. In Sec.II we derive the spin
density induced by a two-dimensional layer of precessing spins,
and the corresponding reaction field. Sec.III focuses on the
dissipative part of the reaction field, and the excess damping
$G'$ for an isotropic electron gas. An expression for $G'$ in a
free electron model is derived for $N/F/N$ structure with $N$
layers of both infinite and finite thickness. The enhancement of
$G'$ due to electron-electron interactions is considered within
the generalized Hartree-Fock approximation. In Sec.IV we establish
a relation between the spin pumping theory of Ref.[11] and our
free electron result for $G'$. The role of spin sink in theories
of Gilbert damping is discussed in Sec.V. The reaction field for
the $N/F/N$ structure is derived in Appendix A. The role of spin
relaxation in the theory of $G'$ is considered in Appendix B.
\section{Dynamic RKKY}

We consider a two-dimensional layer of aligned spins imbedded in
normal metal.  Notice that a similar model was used by Yafet [15]
to calculate the interlayer coupling for time-independent
magnetizations.  Here we take into account the time-dependence of
the precessing magnetization.

Our task is to derive the conduction electron spin density induced
by the $s-d$ exchange interaction taking place at the magnetic
layer.  The Hamiltonian of the conduction electrons is

\begin{equation}\label{Eq5}
\hat{H}=\hat{H}_{0} - J \sum^{sheet}_{i}\int d
^{3}r\vect{S}^{(i)}(t).\hat \vect {s} \vect{(r)}\delta^{3}
(\vect{r}-\vect{r}_{i})
\end{equation}
where $\hat{H}_0$ represents the Hamiltonian of the electrons in
the absence of the ferromagnetic layer, and the second term is the
$s-d$ exchange interaction with the layer.

$J$ is the $s-d$ exchange coupling constant, $\vect{S}^{(i)} (t)$
is the classical spin at the location $\vect{r}_{i}$, and $\hat
\vect {s}(\vect {r})$ is the spin density operator for the
electrons in normal metal.

For a uniform precession of aligned spins, we have
\begin{equation}\label{Eq6}
\vect{S}^{(i)} (t) = \frac{\Omega}{\gamma}\vect{M} (t)
\end{equation}

where $\Omega$ is the volume of the unit cell.

The second term of Eq. (5) acts as a time-dependent perturbation
that will be treated using the linear response theory [16].  Thus,
the expectation value of the $\mu$-component of the induced spin
density is
\begin{eqnarray}
 \langle s_{\mu}(\vect{r},t)\rangle = J \sum^{sheet}_{i}\int
d^{3}r'\int
^{\infty}_{-\infty}dt'\chi_{\mu\nu}(\vect{r}t,\vect{r}'t')
\nonumber \\* \times S^{(i)}_{\nu}(t')
\delta^{3}(\vect{r}'-\vect{r}_{i})
\end{eqnarray}

where
\begin{equation}\label{Eq8}
\chi_{\mu \nu}(\vect{r}t,\vect{r}'t') =
\frac{i}{\hbar}\Theta(t-t')\langle[\hat{s}_{\mu}(\vect{r},t),\hat{s}_{\nu}(\vect{r}',t')]\rangle
\end{equation}
is the retarded spin correlation function (susceptibility)[16],
$\Theta(t-t')$ being the unit step function.

Now, we evaluate the R.H.S. of Eq.(7) assuming a slow precession.
This approximation is valid as long as the precessional frequency
is small compared to the relevant excitation frequencies of the
conduction electrons.  We note that the spin pump theory requires
a similar condition to hold [11, 12].

Performing the $\vect{r}'$-integration, and applying the
commutation law for the convolution, Eq.(7) reads

\begin{equation}\label{Eq9}
\langle
s_{\mu}(\vect{r},t)\rangle=J\sum^{sheet}_{i}\int^{\infty}_{-\infty}dt'S^{(i)}_{\nu}(t-t')\chi_{\mu
\nu}(\vect{r},\vect{r}_{i},t')
\end{equation}

For slow precession, we write

\begin{equation}\label{Eq10}
S^{(i)}_{\nu}(t-t')\simeq
S^{(i)}_{\nu}(t)-t'\frac{dS^{(i)}_{\nu}(t)}{dt}
\end{equation}

Introducing this expansion into Eq.(9), we obtain with use of
Eq.(6)

\begin{eqnarray}\label{Eq11}
\langle s_{\mu}(\vect {r},t)\rangle \simeq
\frac{J\Omega}{\gamma}\lim_{\omega\rightarrow
0}\biggr{[}M_{\nu}(t)\sum^{sheet}_{i}\chi_{\mu
\nu}(\vect{r},\vect{r}_{i};\omega) \nonumber \\* -\frac{\partial
M_{\nu}(t)}{\partial {t}}\sum^{sheet}_{i} \frac{\partial Im
\chi_{\mu \nu}(\vect{r},\vect{r}_{i}, \omega)}{\partial
\omega}\biggr{]}
\end{eqnarray}
where

\begin{equation}\label{Eq12}
\chi_{\mu \nu} (\vect{r},
\vect{r}_{i},\omega)=\int^{\infty}_{-\infty}dt e^{i\omega
t}\chi_{\mu \nu}(\vect{r}, \vect{r}_{i},t)
\end{equation}

We note that the second term on the R.H.S. of Eq.(11) follows by
letting the derivative of the real part of the susceptibility with
respect to $\omega$ equal to zero at $\omega =0$. This is
consistent with the reality condition implying that that the real
part of the susceptibility is an even function of $\omega$. The
first term on the R.H.S. of Eq.(11) corresponds to the static RKKY
result of Yafet [15]. The second term is the dynamic
generalization of RKKY for a slowly precessing magnetization.

For an infinite medium, the sheet sums in eq. (11) can be
evaluated by invoking translational invariance and Fourier
transforming $\chi_{\mu \nu}(\vect{r},\vect{r}_{i},\omega)$ to
$\vect{q}$-space [13]. Thus, we define a generic sum
\begin{eqnarray}\label{Eq13}
X_{\mu \nu}(\vect{r},\omega)=\sum^{sheet}_{i}\chi_{\mu
\nu}(\vect{r},\vect{r}_{i},\omega)=\int
\frac{d^{3}q}{(2\pi)^{3}}\nonumber \\* \times \sum^{sheet}_{i}
\exp[i
\vect{q}.(\vect{r}-\vect{r}_{i})]\chi_{\mu\nu}(\vect{q},\omega)
\end{eqnarray}

from which both terms of Eq.(11) can be deduced.

To perform the $q$-integral on the R.H.S. of this equation, we set

\begin{equation}\label{eq14}
\vect{q}=\vect{q}_{\parallel}+\vect{q}_\perp
\end{equation}

where $\vect{q}_\parallel$ is confined to the sheet, and
$\vect{q}_\perp$ is perpendicular to the sheet.  For a square
sheet of area $L^{2}$, we have
\begin{equation}\label{Eq15}
\int\frac{d^{3}q}{(2\pi)^{3}}=\int\frac{d^{2}q_{\parallel}}{(2\pi)^{2}}\int\frac{d
q_{\perp}}{2\pi}=\frac{1}{L^{2}}\sum_{{q}_{\parallel}}\int\frac{d
q_{\perp}}{2\pi}
\end{equation}

Assuming a continuous distribution of spins, the sheet sum in Eq.
(13) is given by
\begin{equation}\label{Eq16}
\sum^{sheet}_{i} \exp
\bigr{[}-i(\vect{q}_{\parallel}+\vect{q}_{\perp}).\vect{r}_{i}\bigr{]}
=N_{s}\delta_{\vect{q}_{\parallel},0}
\end{equation}

where $N_{s}$ is number of spins in the sheet. Using Eqs. (14-16),
the R.H.S. of Eq.(13) is evaluated with the result

\begin{eqnarray}\label{Eq17}
X_{\mu\nu}(\vect{r},\omega)=\frac{N_{s}}{L^{2}}\int
\frac{dq_{\perp}}{2\pi}
\exp(i\vect{q}_{\perp}.\vect{r})\chi_{\mu\nu}(\vect{q}_{\perp},\omega)
\\*\nonumber =n_{s}\int\frac{dq_{\perp}}{2\pi}\exp
(iq_{\perp}x)\chi_{\mu\nu}( q_{\perp},\omega)
\end{eqnarray}

where $n_{s}=N_{s}/L^{2}$ is the sheet density.  Owing to the
symmetry of the model, the quantity $X_{\mu\nu}(\vect{r},\omega)$
depends only on the distance $x$ from the sheet.

Also, the induced spin density is a function of $x$.  Using
Eqs.(11), (13) and (17), we obtain
\begin{eqnarray}\label{Eq18}
\langle s_{\mu}(x,t)\rangle = \frac{J\Omega}{\gamma}
\lim_{\omega\rightarrow 0}\biggr{[}X_{\mu\nu}(x,\omega)M_{\nu}(t)
\\* \nonumber - \frac{\partial I m X_{\mu\nu}(x,\omega)}{\partial\omega}
\frac{d M_{\nu}(t)}{dt}\biggr{]}
\end{eqnarray}

Using this result and Eq.(17) in Eq.(3), the $\mu$-component of
the reaction field is

\begin{eqnarray}\label{Eq19}
H_{r,\mu}(t)\simeq\frac{2J^{2}\Omega
an_{s}}{\gamma^{2}d}\lim_{\omega\rightarrow 0}\biggr{[} \int
^{\infty}_{-\infty}\frac{d
q}{2\pi}\chi_{\mu\nu}(q,\omega)M_{\nu}(t)
\\*\nonumber
-\frac{\partial}{\partial\omega}\int^{\infty}_{-\infty}\frac{d
q}{2 \pi} I m \chi_{\mu\nu}(q,\omega)\frac{d
M_{\nu}(t)}{dt}\biggr{]}
\end{eqnarray}
The first term on the R.H.S of this equation contributes to the
torque $[\vect{M}\times \vect{H}_{eff}]$ only if $\chi_{\mu\nu}$
is anisotropic.  In this case an anisotropic shift of the FMR
frequency ensues.  On the other hand, the second term contributes
to the dissipative torque that is non vanishing for both isotropic
and anisotropic susceptibility.

\section{gilbert damping}In what follows, we consider the
dissipative torque for an isotropic electron gas. Thus,
$\chi_{\mu\nu}(\vect{q},\omega)=\chi(\vect{q},\omega)\delta_{\mu\nu}$,
and the dissipative part of the reaction field
$\vect{H}_{r}^{(d)}$ is according to Eq.(19) given by
\begin{eqnarray}\label{Eq20}
\vect{H}_{r}^{(d)}(t)\simeq-\frac{2J^{2}\Omega a
n_{s}}{\gamma^{2}d} \\* \nonumber \times \lim_{\omega\rightarrow
0}\biggr{[}
\frac{\partial}{\partial\omega}\int^{\infty}_{-\infty}\frac{dq}{2\pi}I
m\chi(q,\omega)\biggr{]}\frac{d\vect{M}(t)}{dt}
\end{eqnarray}

Introducing this result in the R.H.S. of Eq.(2), we obtain the
damping enhancement constant $G'$
\begin{equation}\label{Eq21}
G'\simeq2J^{2}\Omega
n_{s}M^{2}_{s}\bigr{(}\frac{a}{d}\bigr{)}\lim_{\omega\rightarrow
0}\biggr{[}\frac{\partial}{\partial\omega}\int^{\infty}_{-\infty}\frac{dq}{2\pi}I
m \chi(q,\omega)\biggr{]}
\end{equation}

\subsection {Independent electrons}

First, we evaluate the expression (21) by disregarding the
electron- electron interaction in the $N$ regions.  However, a
finite splitting $\Delta$ of the $\uparrow$ and $\downarrow$ spin
bands is assumed.  The external magnetic field of the FMR
experiment is one source of this splitting. For a system of
infinite size, this splitting establishes a lower cutoff on the
wave vector $q$. As shown below, this cutoff is essential to
prevent the logarithmic divergence of Eq.(21). Due to the spin
splitting, the susceptibility develops some anisotropy. Since
transverse components of the reaction field (19) contribute to the
dissipative torque, we need to use in Eq.(21) the transverse
susceptibility [16]
\begin{equation}\label{Eq22}
\chi^{(0)}_{T}(\vect{q},\omega)=\frac{\hbar^{2}}{4}\int\frac{d^{3}k}{(2\pi)^{3}}
\frac{f_{k+q\downarrow}-f_{k\uparrow}}{\hbar\omega-\epsilon_{k+q\downarrow}
+\epsilon_{k\uparrow}+i\eta}
\end{equation}
where
\begin{equation}\label{Eq23}
\epsilon_{k+q\downarrow}-\epsilon_{k\uparrow}
\simeq\frac{\hbar^{2}\vect{k}.\vect{q}} {{m}}+\Delta
\end{equation}
Expanding the Fermi functions, we have
\begin{equation}\label{Eq24}
f_{k+q\downarrow}-f_{k\uparrow}\simeq\frac{\partial
f}{\partial\epsilon_{k}}\biggr{(}\frac{\hbar^{2}\vect{k}.\vect{q}}{m}+\Delta\biggr{)}
\end{equation}
Using Eqs.(23) and (24), the imaginary part of Eq.(22) becomes
\begin{eqnarray}\label{Eq.25}
I m
\chi^{(0)}_{T}(\vect{q},\omega)\simeq\frac{-\hbar^{2}}{16\pi}\int^{\infty}_{0}
dk k^2 \frac{\partial f}{\partial\epsilon_{k}}\int^{\pi}_{0} d
\theta \sin\theta
\\* \nonumber
\times\biggr{(}\Delta+\frac{\hbar^{2}\vect{k}.\vect{q}}{m}\biggr{)}
\delta\biggr{(}\hbar\omega-\Delta-\frac{\hbar^{2}\vect{k}.\vect{q}}{m}\biggr{)}
\end{eqnarray}
where we used polar coordinates to perform the
$\vect{k}$-integration.  Performing first the integration over
polar angle $\theta$, we have
\begin{eqnarray}\label{Eq.26}
\int^{\pi}_{0}d\theta \sin \theta
\biggr{(}\Delta+\frac{\hbar^{2}\vect{k}.\vect{q}}{m}\biggr{)}
\delta\biggr{(}\hbar\omega-\Delta-\frac{\hbar^{2}\vect{k}.\vect{q}}{m}\biggr{)}
\\* \nonumber=\frac{m\omega}{\hbar k
q}\Theta\biggr{(}k-\frac{|\hbar\omega-\Delta|m}{\hbar^{2}q}\biggr{)}
\end{eqnarray}

Introducing this result into Eq.(25), and converting the
$k$-integral to $\epsilon_{k}$-integration, we obtain at $T=0$

\begin{eqnarray}\label{Eq27}
Im \chi^{(0)}_{T}(q,\omega)=\frac{m^{2}\omega}{16\pi\hbar q
}\int^{\infty}_{0} d\epsilon
\delta(\epsilon-\epsilon_{F})\\*\nonumber \times \Theta
\biggr{(}\sqrt{\frac{2m\epsilon}{\hbar^{2}}} -
\frac{|\hbar\omega-\Delta|m}{\hbar^{2}q}\biggr{)}\\*\nonumber =
\frac {m^{2}\omega}{16\pi\hbar q}\Theta
\biggr{(}k_{F}-\frac{|\hbar\omega-\Delta|m}{\hbar^{2}q}\biggr{)}
\end{eqnarray}

The unit step function on the R.H.S. of this equation equals one
for $q>q_{1}$, and zero for $q<q_{1}$ where

\begin{equation}\label{Eq28}
q_{1}=\frac{|\hbar\omega-\Delta|m}{\hbar^{2}k_{F}}
\end{equation}

Thus, $q_{1}$ acts as a lower cutoff in the $q$-integral of $Im
\chi (q,\omega)$.  Since $|\hbar\omega-\Delta|\ll\epsilon_{F}$,
the upper cutoff is given by $q_{2}\approx2k_{F}$.  We then get
using Eq.(27)

\begin{eqnarray}\label{Eq29}
\int^{\infty}_{-\infty}\frac{dq}{2\pi}I m
\chi^{(0)}_{T}(\vect{q},\omega)\approx\frac{m^{2}\omega}{16\pi^{2}\hbar}
\\* \nonumber\times \int^{q_{2}}_{q_{1}}\frac{dq}{q}=\frac{m^{2}\omega}{16\pi^{2}\hbar}
\ln \frac{4\epsilon_{F}}{|\hbar\omega-\Delta|}
\end{eqnarray}

Introducing this result into Eq.(21), we obtain
\begin{equation}\label{Eq30}
G'\simeq\frac{(J M_{s}a m)^{2}}{8\pi^{2}\hbar d} \ln
\frac{4\epsilon_{F}}{\Delta}
\end{equation}

where we assumed a cubic lattice to write $\Omega n_{s}a=a^{2}$.
Strictly speaking, $G'$ in this equation is not a simple Gilbert
damping since its strength depends on the applied magnetic field,
but it is a weak dependence.

 When $\Delta $ vanishes, $G'$ exhibits a logarithmic divergence.
This is due to the fact that precessing magnetization radiates
into the $N$-region a very large number of electron-hole pairs of
low energy. This is reminiscent of the infrared divergence in
quantum electrodynamics where a large number of low energy photons
is radiated. Endowing the photon with a small mass cures this
divergence [19]. In an analogous way, a nonzero spin splitting of
the electron energy levels suppresses the number of low energy
electron-hole pairs that are generated by the precessing
magnetization.

 For $\Delta\simeq g\mu_{\beta}H$, we have
\begin{equation}\label{Eq31}
\ln\frac{4\epsilon_{F}}{\Delta}=\ln\biggr{(}\frac{2k^{2}_{F}\hbar
c}{eH}\biggr{)}\approx \ln\frac{l_{H}}{a}
\end{equation}
We see that it is the magnetic length
\begin{equation}\label{Eq32}
l_{H}=\biggr{(}\frac{\hbar c}{eH}\biggr{)}^{\frac{1}{2}}
\end{equation}

which plays the role of $q_1^{-1}$ in case of infinitely wide $N$
layers. Taking $H=10^{4}$ gauss and putting $k_{F}\simeq 1/a$, we
obtain from Eq.(32) $l_{H} \simeq 4\times10^{2}a$. In real
samples, the $N$ layers are usually considerably thinner than this
length. As shown in the next subsection, the relevant cutoff is
then determined by the boundary conditions (b.c.) at the outer
surfaces of the sandwich.

It is interesting that spin relaxation of the conduction electrons
can provide an effective gap and a finite cutoff length even in
the limit of infinitely thick $N$ layers. As shown in Appendix B,
the resulting cutoff length is $q_1^{-1}\approx v_F\tau_s $ where
$\tau_s$ is the spin-relaxation time. The corresponding value of
the effective gap is $\Delta _{eff} \approx \hbar /\tau_s$. The
same quantity determines the linewidth of the electron
paramagnetic resonance of conduction electrons [23]. Its magnitude
is presumably not negligible in comparison with the magnetic
splitting $g\mu_B H$.

\subsection{Finite normal layers}
We consider a ferromagnetic sheet imbedded between two $N$ layers
each of thickness $D$. Strictly speaking, in a finite system, the
translational invariance invoked in the evaluation of the sheet
sum (13) is broken. This complicates the calculation of the
induced spin density. Nevertheless, a simplified approximate
evaluation of $X_{\mu\nu}(x,\omega)$ can be carried out if
$x\simeq 0$. In this case, the translational invariance is
restored locally since the boundary at $x=D$ plays a small role.
The b.c. at $x= D$ is taken in the form
\begin{equation}\label{Eq33}
\langle s_{\mu}(x,t)\rangle|_{x=D}=0
\end{equation}
This condition follows from assuming that there is an infinite
potential step at the normal metal-vacuum boundary. Thus, the
$N$-electron wave functions are forced to zero at the boundary and
so is the magnetization density.  Applying this b.c.to Eq.(18),
the finite-$D$ version of Eq.(17) reads
\begin{equation}\label{Eq34}
X_{\mu\nu}(x,\omega) \simeq
\frac{n_s}{D}\sum^{n_{max}}_{n=1}\cos(q_{n}x)\chi_{\mu\nu}(q_{n},\omega)
\end{equation}

where $q_{n}=\pi (n-1/2)/D$, and $n_{max}\simeq D/a$ since it is
the lattice spacing which determines the highest value of $q_{n}$.
In view of this, Eq.(21) is changed to
\begin{equation}\label{Eq35}
G'\simeq \frac{2a}{dD}J^{2}\Omega n_{s}
M^{2}_{s}\lim_{\omega\rightarrow
0}\biggr{[}\frac{\partial}{\partial\omega}\sum
^{n_{max}}_{n=1}Im\chi(q_{n},\omega)\biggr{]}
\end{equation}
For noninteracting electrons without spin splitting, we have
\begin{equation}\label{Eq36}
\sum^{n_{max}}_{n=1}Im\chi(q_{n},\omega)\simeq\frac{m^2\omega}{16\pi
\hbar}\sum^{n_{max}}_{n=1}q_{n}^{-1}
\end{equation}
For $D/a\gg 1$, we use the definition of the Euler constant
$\gamma _E\simeq 1.78$ to obtain

\begin{equation}\label{Eq37}
\sum^{n_{max}}_{n=1} q_{n}^{-1}\approx \frac{D}{\pi}\ln(4\gamma_E
n_{max})
\end{equation}

Using Eqs.(35-37), we obtain

\begin{equation}\label{Eq 38}
G'\approx \frac{(JM_s am)^2}{8\pi^2\hbar d}\ln(D/a)
\end{equation}

As expected, the boundary conditions in a finite slab imply a
cutoff $q_1\approx D^{-1}$.

We now make an order of magnitude estimate of Eq.(38) for an iron
film of thickness $d=D=10a$ where $a=4\times 10^{-8}cm$. The
constant $J$ can be estimated by relating it to the atomic
exchange integral $J_{sd}$
\begin {equation}\label{Eq39}
J\approx \frac{2J_{sd}\Omega}{\hbar^{2}}
\end{equation}

Taking $J_{sd}=0.1 eV$, and $M_s =1.7\times 10^3 gauss$, Eq.(38)
yields $G'\approx 10^8 s^{-1}$. This agrees with the interface
damping observed recently in the double layer structure by Urban
et al.[9]. It should be pointed out that the second ferromagnetic
layer in this experiment plays a crucial role in establishing the
spin sink needed to prevent spin accumulation in the $N$ layers(
see Sec.V.)

\subsection{Electron-electron interactions}

We now calculate $G'$ by taking into account interactions between
electrons in the normal metal. The generalized Hartree-Fock
approximation for the Hubbard model yields the following
expression for the transverse susceptibility [20]
\begin{equation}\label{Eq40}
\chi_{T}(q,\omega)=\frac{\chi_{T}^{(0)}(q,\omega)}{1-\tilde{U}\chi_{T}^{(0)}(q,\omega)}
\end{equation}
where $\tilde{U}=4\Omega U/\hbar^2$, and $U$ is the screened
intraatomic Coulomb energy. Using this formula, we have
\begin{eqnarray}\label{Eq41}
\lim_{\omega\rightarrow 0}\frac{\partial}{\partial\omega}
\biggr{[}Im\chi_{T}(q,\omega)\biggr{]}=\biggr{[}1-\tilde{U}\chi_{T}^{(0)}(q,0)\biggr{]}^{-2}
\\*\nonumber\times \lim_{\omega\rightarrow
0}\frac{\partial}{\partial\omega}\biggr{[}Im\chi_{T}^{(0)}(q,\omega)\biggr{]}
\end{eqnarray}

To simplify the evaluation of the $q$-integral of this quantity,
we take advantage of the weak dependence of the static
susceptibility on $q$ for $q<q_F$, and make the approximation
\begin {equation}\label{Eq42}
\chi^{(0)}_T(q,0)\approx\chi^{(0)}_T(0,0)=\frac{\hbar^2}{4\Omega}N(\epsilon_F)
\end{equation}
where $N(\epsilon_F)$ is the density of states, per atom, at the
Fermi energy.
 Using Eqs.(41) and (42) in Eq.(35), we obtain the
enhancement, $G'$, for interacting electrons in finite $N$ layers

\begin{equation}\label{Eq43}
G'\approx\frac{(JM_{s}amS_E)^{2}}{8\pi^{2}\hbar d}\ln(D/a)
\end{equation}
where $S_E$ is the Stoner factor defined in Eq.(4).  For
palladium, we have $S_E\approx10$.  Thus, large values of $G'$ are
expected for sandwiches containing $Pd$ as normal layers. Mizukami
et al.[8] measured the Gilbert damping constant in $N/F/N$
sandwiches with $F$ being a thing film of permalloy $(Py)$.  These
measurements show that $G'$ for $Pd/Py/Pd$ system is well above
that for the $Cu/Py/Cu$ system.  However, it is about twice as
smaller than $G'$ for $Pt/Py/Pt$.  This seems to contradict our
Eq.(43).  We attribute this disagreement to spin accumulation in
the normal layers.  Since the spin-orbit coupling constant of $Pt$
is larger (by a factor of 3) than that for $Pd$, the spin lattice
relaxation rate in $Pt$ is an order of magnitude stronger than
that in $Pd$. As pointed out by Tserkovnyak et al [11], the spin
accumulation takes place when the spin relaxation rate is small.
The fact that experimental ration $G'_{pt}/G'_{Pd}$ is less than
10 indicates that Stoner enhancement in $Pd$ is not excluded.

More convincing evidence of Stoner enhancement in interface
damping comes from recent FMR studies on
$20Au/40Fe/40Au/3Pd/[Fe/Pd]_{5}/14Fe/GaAs(001)$\\samples [18].
Compared to a single layer structure, these samples show a $G'$
that is enhanced by a factor of four and exhibits a strong
fourfold in-plane anisotropy.  Apparently, the presence of a
second $Fe$ layer provides an efficient spin sink.  Thus $G'$ is
determined by the exchange enhanced susceptibility rather than the
bottleneck due to a weak spin-lattice relaxation in the
$N$-layers.

We now digress, for a moment, to consider the enhancement of the
FMR frequency shift due to interactions. Applying Eq.(40) to the
static anisotropic susceptibility, and making the approximation
(42), we have $\chi_{\mu\nu}(q,0)\approx
S_E\chi_{\mu\nu}^{(0)}(q,0)$. Using this result in Eq.(19), we see
that the frequency shift for the interacting electrons is $S_E$
times that for the independent electrons. This prediction could be
used, in conjunction with the data for the anisotropic $G'$ to
clarify experimentally the role of interactions in the FMR of
multilayers.
\section{Relation to spin-pumping theory}
We now show that for free electrons there is a similarity between
our formula (30) for $G'$ and the spin-pumping theory of
Tserkovnyak et al. [11].  According to these authors, the excess
damping produced by pumping of spins into adjacent N-layers is
$G'=\gamma M_{s}\alpha'$ where
\begin{equation}\label{Eq44}
\alpha'=\frac{g_L\mu_{B}(A_r^{(L)}+A_{r}^{(R)})}{4\pi M_{s}L^{2}d}
\end{equation}

where $g_{L}$ is the Land\'{e} factor, $\mu_{B}$ is the Bohr
magneton, and $A_{r}^{(L)}$, $A_{r}^{(R)}$ are the interface
parameters for the left, right $N$-Layers, respectively.  In terms
of the elements of the $2\times2$ scattering matrix, for a
symmetric $N/F/N$ sandwich, these parameters are

\begin{eqnarray}
A_r^{(L)}=A_r^{(R)}=A_r\\*\nonumber=\frac{1}{2}\sum_{mn}\{|r^{\uparrow}_{mn}-r^{\downarrow}_{mn}|^{2}+|t^{'\uparrow}
_{mn}-t^{'\downarrow}_{mn}|^{2}\}
\end{eqnarray}

where $(r^{\uparrow}_{mn},r^{\downarrow}_{mn})$ and
$(t^{\uparrow}_{mn}, t^{\downarrow}_{mn})$ are the reflection and
transmission coefficients for electrons with up and down spins.
The expression (45) is to be evaluated with the transverse modes
$(m,n)$ taken at the Fermi energy.

Following Bruno [13], we consider the scattering of $N$-electrons
by a ferromagnetic monolayer.  Due to the conservation of
transverse momentum, $r_{mn}=r_{m}\delta_{mn}$ where
$r_{m}=r_0(k_{\perp})$, $k_{\perp}$ being the component of the
electron wave vector perpendicular to the monolayer.

The reflection coefficients $r_0(k_{\perp})$ are found by solving
the one-dimensional scattering problem for the potential
\begin{equation}\label{Eq46}
v(x)=v_0\delta(x)
\end{equation}
where $v_0$ is given by the interface coupling constant $J$, and
by the magnitude of the atomic spin $S$ of the ferromagnet
\begin{equation}\label{Eq47}
v_0=\pm\frac{\hbar}{2}JSn_s
\end{equation}
where the $(+,-)$ signs correspond to the $(\downarrow,\uparrow)$
electron spins, respectively.

The reflection coefficients for this problem are [13]
\begin{eqnarray}\label{Eq48}
r_0^{\uparrow}=\frac{-i\beta}{k_{\perp}+i\beta}\\*\nonumber
r_0^{\downarrow}=\frac{i\beta}{k_{\perp}-i\beta}
\end{eqnarray}
where
\begin{equation}\label{Eq49}
\beta=\frac{mv_0}{\hbar^{2}}
\end{equation}
The transmission coefficients are [20]
\begin{eqnarray}\label{Eq50}
t_0^{\uparrow}=\frac{ik_{\perp}}{ik_{\perp}-\beta}\\*\nonumber
t_0^{\downarrow}=\frac{ik_{\perp}}{ik_{\perp}+\beta}
\end{eqnarray}

Eqs.(48) and (50) imply
\begin{eqnarray}\label{Eq51}
\bigr{|}r_0^{\uparrow}-r_0^{\downarrow}\bigr{|}^2_{\epsilon_F}
=\bigr{|}t_0^{\uparrow}-t_0^{\downarrow}\bigr{|}^2_{\epsilon_F}\\*\nonumber
=\frac{4{\beta}
^2(k_F^2-k_{\parallel}^2)}{(k_F^2-k_{\parallel}^2+\beta^2)^2}
\end{eqnarray}
where we applied the identity
$(k_{\parallel}^2+k_{\perp}^2)_{\epsilon_F}=k_F^2$. Using this
result, the transverse mode sum (45) becomes a sum over the
in-plane wave vectors $\vect k_{\parallel}$. Converting the $\vect
k_{\parallel}$-sum to a two-dimensional integral, we have
\begin{equation}\label{Eq52}
A_r\simeq\frac{L^2}{2\pi}\int_{0}^{k_F}dk_{\parallel}
k_{\parallel}
\frac{4\beta^2(k_F^2-k_{\parallel}^2)}{(k_F^2-k_{\parallel}^2+\beta^2)^2}
\end {equation}
Evaluating this integral, we get
$A_r\simeq(L^2\beta^2/\pi)F(\beta)$ where
\begin{equation}\label{Eq53}
F(\beta)\simeq
\ln\frac{k_F^2+\beta^2}{\beta^2}-\frac{k_F^2}{k_F^2+\beta^2}
\end{equation}
Inserting Eqs.(52) and (53) into Eq.(44) and expressing $\beta$
with use of Eqs.(47) and (48), we obtain
\begin{equation}\label{Eq54}
\alpha'\simeq\frac{\gamma(mSJn_s)^2}{8\pi^2\hbar M_sd}F(\beta)
\end{equation}
We can bring this result closer to the form of Eq.(30) by letting
$n_s=a^{-2}$, and $M_s=\gamma Sa^{-3}$ yielding
\begin{equation}\label{Eq55}
\gamma S^2n_s^2=\frac{M_s^2a^2}{\gamma}
\end{equation}
Furthermore, assuming $\beta\ll k_F$, the function $F(\beta)$ can
be approximated by
\begin{equation}\label{Eq56}
F(\beta)\approx 2\ln\frac{k_F}{1.65\beta}\approx
2\ln\frac{\epsilon_F}{J_{sd}}
\end{equation}
>From Eqs.(54-56) we have
\begin{equation}\label{Eq57}
G'=\gamma M_s \alpha'\approx \frac{(JM_sam)^2}{4\pi^2 \hbar
d}\ln\frac{\epsilon_F}{J_{sd}}
\end{equation}
This equation shows a remarkable similarity with the expression
(30). Note, however, that the logarithmic terms do not match. If
we consider an infinite system, and ignore the cutoff due to the
magnetic length (32), the gap $\Delta$ vanishes , and Eq.(30)
becomes logarithmically divergent. On the other hand, Eq.(57)
shows that there is an effective gap, $\Delta\approx J_{sd}$
corresponding to a finite cutoff $q_1\approx
k_F(J_{sd}/\epsilon_F)$. Thus, the spin-pumping theory is
infrared-divergence free. For real finite size systems, this
difference is only of academic interest since the cutoff produced
by the boundary conditions, $q_1\approx D^{-1}$, yields a
logarithmic term that is of the same order of magnitude as that in
eq.(57).

The presence of an effective gap, $\Delta\approx J_{sd}$, in the
spin pumping theory is presumably linked to the fact that, in
contrast to linear response theory, it is of infinite order in the
coupling constant $J$.  This is seen in Eq.(54) where $F(\beta) $
is a nonlinear function of $\beta$ given in Eq.(53). This kind of
nonlinearity is generic in the scattering approach to
transport(see Eq.(51)). In fact, Bruno [13] derives an exact
expression for the static RKKY coupling that goes beyond the
linear response result of Yafet [15].
\section{DISCUSSION}
Our numerical estimate of $G'$ based on Eq.(38) suggests that a
substantial enhancement of the FMR linewidth, that is independent
of the atomic number $Z$, should be observed in $N/F/N$ systems.
In contrast, the data of Mizukami et al.[8] on trilayers
containing permalloy films show a strong dependence of $G'$ on
$Z$. In fact, for $Cu$, which has the smallest $Z$ of the
$N$-metals studied there is a complete absence of a
$1/d$-dependent $G'$. Tserkovnyak et al. [11] propose that it is
the spin accumulation in the $N$-layer that is responsible for
such a suppression of the ferromagnetic relaxation in copper
layer. Note that the theory of spin-pumping assumes at the outset
that the spin system in the $N$-layer is kept in thermal
equilibrium during the precession. For that one needs an efficient
spin-sink mechanism. The data of Ref.[8] indicate that
spin-lattice relaxation via spin-orbit coupling [21] provides the
required spin sink. In fact, metals with larger $Z$ exhibit
generally larger measured values of $G'$. This trend is in
agreement with the fact that the spin-lattice relaxation rate
scales as $Z^4$ [22].

Also the theory of $G'$, presented in Sec.III, assumes that the
electron spins in the $N$-layers are in thermal equilibrium. This
can be established either by the spin-lattice relaxation in the
bulk, or by surface relaxation. One way to include these effects
into the reaction field of Eq.(3) is to calculate the quantity
$\vect\langle\vect s(0,t)\rangle$ using the Bloch equation with
diffusion [23]. This equation is to be solved with the b.c. that
allows the electron spin to be flipped upon collision with the
surface. Such b.c. has been proposed by Dyson [23]. In terms of
the spin density $\langle \vect s\rangle$ this so called
"evaporation" b.c. reads
\begin{equation}\label{Eq58}
\frac{\partial \langle\vect s \rangle}{\partial
x}=\frac{3p}{4\Lambda}\langle \vect s \rangle
\end{equation}
where $p$ is the probability of spin flip to take place upon the
reflection from the boundary, and $\Lambda$ is the mean-free path
in the bulk. Due to surface irregularities and paramagnetic
surface impurities, the probability $p$ can be large enough to
provide the necessary spin sink even for layers with small bulk
disorder.

Alternatively, the spin density and the reaction field can be
obtained from the time-dependent $2\times 2$ matrix kinetic
equations driven by precessing magnetization of the ferromagnet
[24]. Such an approach is inspired by the work of Kambersk\'{y}
[25] on intrinsic damping due to spin-orbit coupling in bulk
ferromagnets. This work invokes the idea of a "breathing Fermi
surface": The chemical potential varies in response to the
time-dependent perturbation. However, the distribution of the
electrons does not respond instantaneously to the perturbation.
There is a time-lag characterized by a relaxation time $\tau$. In
Ref.[18] we apply this idea to the case of dynamic interlayer
exchange coupling. In this case it is the spacer electrons which
are affected by the spin-dependent potential at the interfaces,
and the time variation of this potential is due to the precession
of the ferromagnetic moment. Thus, the relevant relaxation time is
the transverse spin-relaxation time $\tau_{spin}$. The resulting
effective damping field, is like Eq.(20), proportional to $d\vect
M(t)/dt$ implying Gilbert damping. However, in distinction from
Eq.(20), it is also proportional to $\tau_{spin}$.

 This brings us to
the question: Is there a system to which the ballistic theory of
the present paper is applicable? We believe that the double-layer
structure studied in Ref.[9] is a good example of such a system.
Here, the precessing layer $F_1$ deposits spin current into the
$N$-spacer and the second layer $F_2$ acts as an absorber of the
transverse component of the spin current - thus providing an
effective spin sink. Detailed analysis of this mechanism has been
presented in beautiful papers by Stiles and Zangwill [26,27].
These authors show that there is an oscillatory, power law, decay
of the transmitted transverse-spin current that is caused by
cancellations due to a distribution of precessional frequencies,
and the rotation of the spin of the incoming spin upon reflection.
Consequently, almost complete cancellation of the transverse spin
takes place after propagation into the ferromagnet by a few
lattice constants. This finding also supports our assumption that
the excitation of transverse components of $\langle \vect
s(x,t)\rangle$, via $s-d$ exchange, is confined to the $N/F$
interface layer(see Appendix A).
\section*{Acknowledgments}
 Research by B.H. is supported by the Natural Sciences and
 Engineering Research Council of Canada (NSERC) and the Canadian
 Institute for Advanced Research (CIAR). E.\v{S}. wishes to express
 his thanks to CIAR for supporting his visit to Simon Fraser
 University.

\appendix
\section{DERIVATION OF EQ.(3)}
We consider a trilayer shown in Fig.1., and derive the reaction
field from the torque equation
\begin{equation}\eqnum{A1}
\biggr{[}\vect M_f(t)\times \vect H_r(t)\biggr{]}=\vect T(t)
\end{equation}

where $\vect M_f(t)$ is the net magnetic moment of the
ferromagnetic film, and $\vect T(t)$ is the torque due to the
$s-d$ interaction. Since only interface regions contribute to this
torque (see Refs.[13] and [27]), we pick a magnetic atom in the
plane $x=0$, and consider the local magnetic field $\vect
H^{(i)}(t)$ acting on its magnetic moment $\vect M^{(i)}(t)$.
Using Eq.(5), the expectation value of the $s-d$ exchange energy
of this atom is $-J\vect S^{(i)}(t).\langle\vect s(0,t)\rangle$.
If we write this quantity as $-\vect M^{(i)}(t).\vect H^{(i)}(t)$,
where $\vect M^{(i)}(t)=\gamma \vect S^{(i)}(t)$, the local
magnetic field is

\begin{equation}\eqnum{A2}
\vect H^{(i)}(t)=\frac{J}{\gamma}\langle\vect s(0,t)\rangle
\end{equation}
 For a square film of area $L^2$, the number of interface atoms is
 $L^2/a^2$. Thus, using Eq.(A2), the net torque contributed by both interfaces is
 \begin{equation}\eqnum{A3}
 \vect T=\frac{2 L^2 J}{a^2\gamma}\biggr{[}\vect M^{(i)}(t)\times \vect
 \langle\vect s(0,t)\rangle\biggr{]}
 \end{equation}
Noting that the net magnetic moment of the film is $\vect
M_f=\vect M^{(i)}L^2d/a^3$, and using Eq.(A3) in (A1), yields
Eq.(3). Similar approach has been used to deduce the effective
field in ultrathin layers in the presence of interfaces (see
Eq.(1.6) in Ref.[3]).

\section{SPIN RELAXATION AND INFRARED CUTOFF}

To include spin relaxation into the theory of $G'$, we start from Eq.(22)
 and replace the infinitesimal quantity $\eta$ by $\Gamma=\hbar/\tau_s$, where $\tau_s$
 is the spin-relaxation time. Thus, the electron-hole pairs with flipped spin are assumed
  to relax with a frequency $\Gamma/\hbar$. Similar assumption has been made in the theory
   of magnon relaxation via $s-d$ interaction [28]. Moreover, we neglect the splitting
$\Delta$. Making these changes in Eqs.(22)-(24), we obtain
\begin{equation}\eqnum{B1}
Im\chi_T(q,\omega)\approx-\frac{\hbar^2\Gamma}{16\pi^2}
\int_{0}^{\infty}dkk^2\frac{\partial f}{\partial\epsilon_k}I(k,q)
\end{equation}
where
\begin{equation}\eqnum{B2}
I(k,q)=\int_{0}^{\pi}d\theta \sin{\theta}\frac{\hbar^2
kq\cos{\theta}/m}{(\hbar\omega-\hbar^2kq\cos{\theta}/m)^2+\Gamma^2}
\end{equation}

Expanding the integrand to order $\omega$, the
$\theta$-integration yields
\begin{equation}\eqnum {B3}
I(k,q)=\frac{2\hbar\omega}{\Gamma^2x} \biggr{[}
\tan^{-1}x-\frac{x}{1+x^2}\biggr{]}
\end{equation}
where $x=\hbar^2kq/(\Gamma m)$. Using Eq.(B3) , the
$k$-integration in Eq.(B1) yields at $T=0$
\begin{eqnarray}\eqnum{B4}
Im\chi_T(q,\omega)\approx\frac{m^2\omega}{8\pi^2\hbar
q}\biggr{[}\tan^{-1}(v_F\tau_s q)\\*\nonumber
-\frac{v_F\tau_sq}{1+(v_F\tau_s q)^2}\biggr{]}
\end{eqnarray}
Consistent with Eq.(27), this expression is equal to
$m^2\omega/(16\pi\hbar q)$ in the limit $\tau_s\rightarrow
\infty$. The evaluation of the $q$-integral of this expression is
done by approximating $\tan^{-1}x$ by $\pi x/2$ for $x<1$, and by
$\pi /2$ for $x>1$. For $\epsilon_F\gg \Gamma$, we obtain
\begin{equation}\eqnum{B5}
\int_{-q_2}^{q_2}\frac{dq}{2\pi}
Im\chi_T(q,\omega)\approx\frac{m^2\omega}{16\pi^2\hbar}\ln(v_F\tau_sq_2)
\end{equation}

Introducing this result into Eq.(21), the damping enhancement
 in the presence of spin relaxation is given by
 \begin {equation}
 G'\approx\frac{(JM_sam)^2}{8\pi^2\hbar d}\ln\frac{q_2}{q_1}
 \end{equation}
 where $q_1\approx(v_F\tau_s)^{-1}$ is the infrared cutoff mentioned at the end of Sec.IIIA.   

\pagebreak
\begin{figure}
\caption{A trilayer consisting of normal metals($N$) adjacent to a
ferromagnetic film ($F$) of thickness $d$. The $s-d$ interaction
generating the spin density $\vect s$ is assumed to take place in
contact layers of thickness, $a$, of the order of lattice
constant.} \label{Fig.1.}
\end{figure}
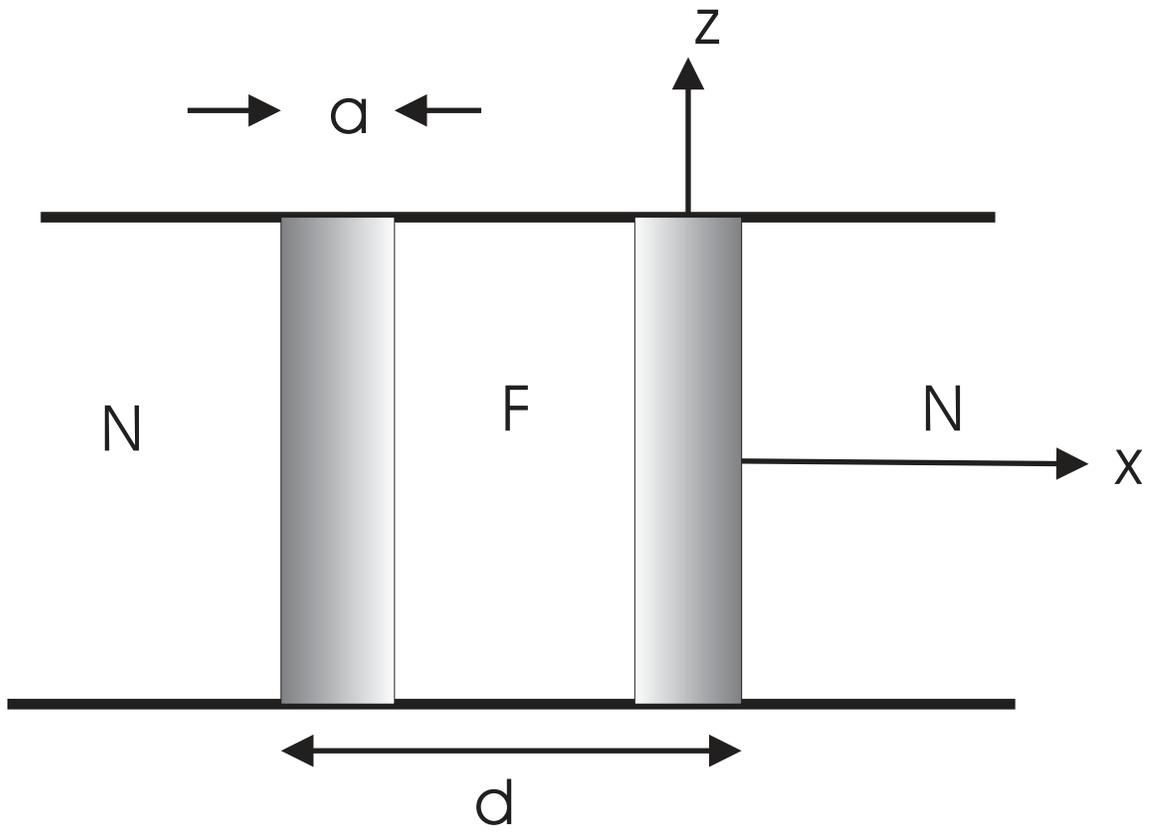
\end{document}